\documentclass[twocolumn,showpacs,preprintnumbers,amsmath,amssymb,floatfix,nofootinbib]{revtex4}

\usepackage{graphicx}
\usepackage{dcolumn}
\usepackage{bm}

\def \pom {{I\!\!P}}

\begin{document}

\title{A study on central diffractive $f_0(980)$ and $f_2(1270)$ meson production at the LHC}

\author{M. V. T. Machado}
\affiliation{High Energy Physics Phenomenology Group, GFPAE  IF-UFRGS \\
Caixa Postal 15051, CEP 91501-970, Porto Alegre, RS, Brazil}

\begin{abstract}
In this contribution we report on the investigation of central diffractive production of mesons $f_0(980)$ and $f_2(1270)$ at the energy of Large Hadron Collider in proton-proton collisions. The cross sections and rapidity distributions for these low-lying mesons are analyzed. The Pomeron-Pomeron process was considered relying on a nonperturbative Pomeron model which is justified by the low mass of considered resonances. The associated theoretical uncertainties are discussed. 
\end{abstract}


\maketitle

\section{Introduction}

In the last two years, ALICE collaboration has recorded zero bias and minimum bias data in proton-proton collisions at a center-of-mass energy of $\sqrt{s}=7$ TeV. Among the relevant events, those containing double gap topology have been studied and they are associated to central diffractive processes \cite{ALICE}. In particular, central meson production was observed. In the double gap distribution, the $K_s^0$ and $\rho^0$ are highly suppressed while the $f_0(980)$ and $f_2(1270)$ with quantum numbers $J^{PC}=(0,2)^{++}$ are much enhanced. Such a measurement of those states is evidence that the double gap condition used by ALICE selects events dominated by double Pomeron exchange. The central exclusive production processes in high energy collisions have traditionally considered a promising way to study particles in a especially clean environment in which to measure the nature and quantum numbers, e.g. spin and parity, of the centrally produced state.  In addition they give information about the structure of the Pomeron and of the mechanism of Pomeron-Pomeron interaction. 

Here, we summarize the main results presented in Ref. \cite{MVTMPRD}, where we focused on the central diffractive production of mesons $f_0(980)$ and $f_2(1270)$ in proton-proton collisions at LHC energies. This investigation is relevant for the ATLAS, CMS and ALICE experiments. We will consider the Pomeron-Pomeron processes and investigate their theoretical uncertainties. In particular, we consider nonperturbative Pomeron model which is justified by the low mass of considered resonances. This contribution is organized as follows: in next  section we present the main expressions for cross section calculation of two-Pomeron processe and in last section we shown the numerical results and discussions. 

\section{Cross section calculation}

Here, we investigate the exclusive meson production from central diffractive reactions. That is the Pomeron-Pomeron production channel. In particular, we focus on the central diffraction (double Pomeron exchange, DPE) in proton-proton interactions. Due to the light mass of mesons $f_0(980)$ and $f_2(1270)$ a non-perturbative approach for the Pomeron exchange should be reasonable. We first focus on the isoscalar meson $f_0(980)[0^{++}]$. We will start with the phenomenological model for the soft Pomeron \cite{Land-Nacht,Bial-Land}.  In the exclusive DPE event the central object $X$ is produced alone, separated from the outgoing hadrons by rapidity gaps,
$pp\rightarrow p+\text{gap}+X+\text{gap}+p$. In approach we are going to use, Pomeron exchange corresponds to the exchange of a
pair of non-perturbative gluons which takes place between a pair of colliding quarks. The scattering matrix is given by,
\begin{eqnarray}
\mathcal{M} & = & \mathcal{M}_{0}\left(  \frac{s}{s_{1}}\right)  ^{\alpha(t_{2})-1}\left(  \frac{s}{s_{2}}\right)
^{\alpha(t_{1})-1}\,F(t_{1})\,F(t_{2})\nonumber \\
&\times & \exp\left(  \beta\left(  t_{1}+t_{2}\right)  \right)\,  S_{\text{gap}}\left(\sqrt{s} \right).
\label{M_all}
\end{eqnarray}
Here $\mathcal{M}_{0}$ is the amplitude in the forward scattering limit ($t_1=t_2=0$). The standard Pomeron Regge
trajectory is given by $\alpha\left(  t\right)=1+\epsilon+\alpha^{\prime}t$ with
$\epsilon\approx 0.08,$ $\alpha^{\prime}=0.25$ GeV$^{-2}$. The momenta of incoming (outgoing) protons are labeled by $p_1$
and $p_2$ ($k_1$ and $k_2$), whereas the glueball momentum is denoted by $P$. Thus, we can define the following quantities
appearing in Eq. (\ref{M_all}): $s=(p_{1}+p_{2})^{2}$, $s_{1}=(k_{1}+P)^{2},$ $s_{2}=(k_{2}+P)^{2},$
$t_{1}=(p_{1}-k_{1})^{2}$ , $t_{2}=(p_{2}-k_{2})^{2}$. The nucleon form-factor is given by $F_p\left(  t\right)  $ = $\exp(b
t)$ with $b=$ $2$ GeV$^{-2}$. The phenomenological factor $\exp\left(  \beta\left(  t_{1}+t_{2}\right)  \right)
$ with $\beta$ $=$ $1$ GeV$^{-2}$ takes into account the effect of the momentum transfer dependence of the non-perturbative
gluon propagator. The factor $S_{\text{gap}}$ takes the gap survival effect into account $i.e.$ the probability
($S_{\text{gap}}^{2}$) of the gaps not to be populated by secondaries produced in the soft rescattering.  For our purpose
here, we will consider $S_{\mathrm{gap}}^2=0.026$ at $\sqrt{s}=14$ TeV in nucleon-nucleon collisions \cite{KKMR}. Such a value for the survival gap factor is typical for soft processes and should reasonable for a cross section estimation involving light mesons as $f_0$ (its mass scale is typical from the nonperturbative regime). However, it should be noticed that the particular values of $S_{\mathrm{gap}}^2$ is dependent on the mass and spin of centrally produced system. An updated discussion on those dependencies for the heavy $\chi_{c,b}$ mesons can be found in Ref. \cite{KMRCHIS2}. When computing cross sections at 7 TeV we will use the interpolation value $S_{\mathrm{gap}}^2=0.032$.

Following the calculation presented in Ref. \cite{Bial-Land} we find
$\mathcal{M}_{0}$ for colliding hadrons,
\begin{eqnarray}
\mathcal{M}_{0}=32 \,\alpha_0^2\,D_{0}^{3}\,\int d^{2}\vec{\kappa}\,p_{1}^{\lambda}V_{\lambda\nu}^{J}p_{2}^{\nu}\,
\exp(-3\,\vec{\kappa }^{2}/\tau^{2}),
\label{M_o}
\end{eqnarray}
where $\kappa $ is the transverse momentum carried by each of the three gluons. $V_{\lambda\nu}^{J}$ is the
$gg\rightarrow R^{J}$ vertex depending on the polarization $J$ of the $R^{J}$ meson state. The fixed parameters of model are $\tau=1$ GeV and $D_{0}G^{2}\tau=30$ GeV$^{-1}$ \cite{Bial-Land} where $G$ is the scale
of the process independent non-perturbative quark gluon coupling.  We consider the parameter
$\alpha_0=G^2/4\pi$ as free and it has been constrained by the experimental result for $\chi_{c}\,(0^{++})$ production at Tevatron \cite{TevCHI}, $d\sigma\,(\chi_{c0})/dy|_{y=0}=76\pm 14 $ nb.  Namely, we found the constraint $S_{\mathrm{gap}}^2\,(\sqrt{s}=2\,\mathrm{TeV})/\alpha_0^2 = 0.45$, where $S_{\mathrm{gap}}^2$ is the gap survival probability factor (absorption factor). Considering the KMR \cite{KKMR} value $S_{\mathrm{gap}}^2=0.045$ for central diffractive processes at Tevatron energy, one obtains $\alpha_0=0.316$. We notice that the CDF collaboration \cite{TevCHI} assumes the absolute dominance of the spin-$0$ contribution in the charmonium production in the radiative $J/\Psi+\gamma$ decay channel (the events had a limited mass resolution and were collected in a restricted area of final-state kinematics)  and then the result was published as:
\begin{eqnarray}
\left.\frac{d\sigma[\chi_c(0^+)]}{dy}\right|_{y=0}& \simeq & \frac{1}{\mathrm{BR}[\chi_c(0^+)]}\left.\frac{d\sigma[pp\rightarrow pp(J/\Psi+\gamma)]}{dy}\right|_{y=0}\nonumber \\
&=& (76\pm 14)\,nb,
\end{eqnarray}
where $\mathrm{BR}[\chi_c(0^+)]=\mathrm{BR}(\chi_c(0^+)\rightarrow J/\Psi+\gamma)$ is the corresponding branching ratio. This fact is not true for general kinematics, as indicated by the Durham group investigations \cite{KMRCHIS2}.

For the
isoscalar meson $f_0(980)$ considered here, $J=0$, one obtains the following result
\cite{Bial-Land,KMRS}:
\begin{equation}
p_{1}^{\lambda}V_{\lambda\nu}^{0}p_{2}^{\nu}=\frac{s\,\vec{\kappa}^{2}
}{2M_{G^{0}}^{2}}A, \label{p_1Vp_2}%
\end{equation}
where $A$ is expressed by the mass $M_{G}$ and the width $\Gamma (gg\rightarrow R)$ of the meson resonance $R$ through the relation:
\begin{equation}
A^{2}= 8\pi M_R\,\Gamma (gg\rightarrow R). \label{A^2}
\end{equation}
For obtaining the two-gluon decays widths the following relation is used,
$\Gamma \,(R\rightarrow gg)=\mathrm{Br}\,(R\rightarrow gg)\,\Gamma_{tot}(R)$. For simplicity, we will take $\mathrm{Br}\,(R\rightarrow gg)=1$, which will introduce a sizable theoretical uncertainty. The two-gluon width depends on the branching fraction of the resonance $R$ to gluons. It is timely to mention that for scalar mesons which are glueballs candidates, is a theoretical expectation \cite{Farrar} that $\mathrm{Br}\,(R(q\bar{q})\rightarrow gg)={\cal O}(\alpha_s^2)\simeq 0.1-0.2$ whereas $\mathrm{Br}\,(R(G)\rightarrow gg)\simeq 1$.  The values for $\Gamma_{gg}$ used in our calculations are summarized in Table I. The numerical results for the LHC energies in proton-proton collisions are presented also in the Table I. The rapidity distribution $d\sigma (y=0)/dy$ and the total integrated cross section are computed. These values can be compared to previous calculations including production in the heavy-ion mode summarized in Ref. \cite{hep_lowx}.

A limitation of the approach above is that it does not allow to deal with $J=1,2$ states.  This is the case for the meson $f_0(1270)$, which is a state $J^{PC}=2^{++}$. It has been shown \cite{Yuan} that the DPE contribution to $J=1$ and $J=2$ meson production in the forward scattering limit is vanishing, either perturbative or non-perturbative Pomeron models. This limitation can be circumvented if we consider Donnachie-Landshoff Pomeron \cite{DOLA}, where it is considered like a isoscalar ($C=+1$) photon when coupling to a quark or anti-quark. In this approach, the cross section is written as:
\begin{eqnarray}
&& \sigma \,(pp\rightarrow p+R+p)  =   \frac{1}{2(4\pi )^3s^2W_{R_J}^2}\int dP_{R} \,dt_1 dt_2 \nonumber \\
 & & \times  \sum_{j=1}^{2}\omega_j\ell_1^{\mu \alpha} \ell_2^{\nu \beta}\,A_{\mu \nu}^JA_{\alpha \beta}^{J*}\left[D_{\pom}(t_1)D_{\pom}(t_2) \right]^2.
\end{eqnarray}

Here, the effective Pomeron propagator is giving by,
\begin{eqnarray}
D_{\pom}(t)=3\beta_0^2\left(\frac{\omega }{E}\right)^{1-\alpha_{\pom}(t)}\,F_p(t),
\end{eqnarray}
where $\beta_0=1.8$ GeV$^{-1}$ and the form factor $F_p(t)$ of the nucleon is taken into account in the form of
\begin{eqnarray}
F_p(t) = \frac{4m_p^2-2.8t}{4m_p^2-t}\left(1-\frac{t}{0.7\mathrm{GeV}^2}\right)^{-2}.
\end{eqnarray}
The coupling to the nucleon is described by the tensor $\ell^{\mu \alpha}$ arising from its fermionic current. For the Pomeron-energies it is used $\omega_{1,2}=(W_{R_J}\pm P_{R})/2$, where $W_{R_J}$ corresponds to the total energy of the meson $R_J$ in the center-of-mass system given by $W_{R_J} = P_{R}^2+M_R^2$. Concerning the Pomeron-Pomeron-R vertex, the $R$ particle is treated as a non-relativistic bound state of a $q\bar{q}$ system. Since the Pomeron couples approximately like a $C=+1$ photon, the Pomeron-quark vertex is given by a $\gamma$-matrix. For the amplitude, we show the explicit formulae:
\begin{eqnarray}
A_{\mu \nu}^{J=0} & = & A_0\left\{ \left[g_{\mu \nu}(q_1\cdot q_2)-q_{2\mu}q_{1\nu} \right]\left[M_R^2+  (q_1\cdot q_2)\right]\right. \nonumber \\
& - & \left. g_{\mu\nu}q_1^2q_2^2 \right\},\nonumber \\
A_{\mu \nu}^{J=1} & =&  A_1\left(q_1^2\epsilon_{\alpha \mu \nu \beta} \epsilon^{\alpha}q_2^{\beta}- q_2^2\epsilon_{\alpha \mu \nu \beta} \epsilon^{\alpha}q_1^{\beta}\right),\nonumber \\
A_{\mu \nu}^{J=2} & = & A_2\left[(q_1\cdot q_2)g_{\mu \rho}g_{\nu \rho}+g_{\mu\nu}q_{1\rho}q_{2\sigma} \right. \nonumber \\
& - & \left. q_{2\mu}q_{1\rho}g_{\sigma\nu}-q_{1\nu}q_{2\rho}g_{\sigma \mu}  \right]\epsilon^{\rho \sigma}, \nonumber
\end{eqnarray}
where $A_0 = \frac{2}{\sqrt{6}}\frac{a}{M_R}$, $A_1 = ia$ and $A_2=\sqrt{2} a M_R$. The formulae above have been firstly obtained in Ref. \cite{KKS} for photon-photon fusion into a quarkonium state. In addition, $\epsilon_{\mu}$ and $\epsilon_{\mu \nu}$ are the polarization vector and tensor of the $J=1$ and $J=2$ states, respectively. The factor $a$ is given by:
\begin{eqnarray}
a=\frac{4}{(q_1\cdot q_2)}\sqrt{\frac{6}{4\pi\,M_R}}\,\phi^{\prime}(0)\,
\end{eqnarray}
where $\phi^{\prime}(0)$ denotes the derivative of the wavefunction at the origin in coordinate space, which can be determined from meson two-photon width  $\Gamma(R_{J=2}\rightarrow \gamma\gamma)$.

We focus now on the estimate for the $f_2(1270)$ meson. The non-relativistic quark model  predicts that its two-photon partial width is given by \cite{Barnes},
\begin{eqnarray}
\Gamma_{\gamma \gamma}(f_2(1270)) = 3\left(\frac{5}{9\sqrt{2}}\right)^2\frac{12}{5}\frac{2^4\alpha^2}{M^4}|\phi^{\prime}(0)|^2.
\end{eqnarray}
Our numerical results for the LHC energy are presented in Table II. The prediction for Tevatron at $\sqrt{s}=1.96$ TeV is $\sigma_{tot}(f_2)=1058 $ nb. We have checked that the result agrees in order of magnitude with the CERN WA102 data \cite{WA102} at $\sqrt{s}=29.1$ GeV (using $S_{\text{gap}}^2=1$). We found $\sigma_{th}=5130$ nb versus the experimental value $\sigma_{exp} = 3275\pm 422$ nb.
We have computed the $f_0(980)$ cross section at $\sqrt{s}=7$ TeV in the approach above, giving $\sigma_{tot}(f_0(980))\simeq 10$ $\mu$b, which is at least one order of magnitude below the soft Pomeron model (see Table I).

\begin{table}[t]
\centering
\renewcommand{\arraystretch}{1.5}
\begin{tabular}{c c c c}
\hline
 $f_0(980)$ & $\Gamma_{tot}$ [MeV] & $\sqrt{s}=7$ TeV & $\sqrt{s}=14$ TeV \\
\hline
  $\frac{d\sigma}{dy}(y=0)$  & ($70\pm 38$)   & 26.9 $\mu$b  & 27.1 $\mu$b \\
  $\sigma_{tot}$  & --- &  369 $\mu$b &  407 $\mu$b \\
  \hline
\end{tabular}
\caption{Rapidity distribution at $y=0$ and integrated cross sections for isoscalar meson $f_0(980)$ in the soft Pomeron model at the LHC energies \cite{MVTMPRD}.}
\label{tab1}
\end{table}

In what follows we discuss the general features and uncertainties in the two referred production channels and in addition we discuss an estimate of $f_2(2170)$ produced in a photon-Odderon channel.

\section{Results and discussions}
\begin{table}[t]
\centering
\renewcommand{\arraystretch}{1.5}
\begin{tabular}{c c c c}
\hline
 $f_2(1270)$ & $\Gamma_{\gamma \gamma}/\Gamma_{tot}$  & $\sqrt{s}=7$ TeV & $\sqrt{s}=14$ TeV \\
\hline
  $\sigma_{tot}$  & $(1.64\pm 0.19)\times 10^{-5}$ & 1083 nb &  1107 nb \\
  \hline
\end{tabular}
\caption{Integrated cross sections for the meson $f_2(1270)$ in the Donnachie-Landshoff Pomeron  model at the LHC energies \cite{MVTMPRD}.}
\label{tab2}
\end{table}

In Table I the cross sections for scalar meson production in the soft Pomeron model  at the LHC energies are shown. The cross sections are reasonably large, despite the nonperturbative Pomeron  energy behavior of the considered model. The deviation is quite sizable when considering the Donnachie-Landshoff Pomeron, which gives  $\sigma_{f_0}(\sqrt{s}=7 \,\mathrm{TeV}) \approx 10$ $\mu$b. The deviation by a factor ten remains even accounting for a small branching ratio of scalar meson in two-gluons (we have used the simplification $\Gamma_{gg}=\Gamma_{tot}$). The predictions for meson $f_2(1270)$ considering the Donnachie-Landshoff Pomeron at the LHC energy are presented in Table II, whereas we found for Tevatron the estimate $\sigma_{f_2}=1058 $ nb.

As a final analysis, we would like to address the photon-Odderon production channel in the specific case of the meson $f_2(1270)$. The phenomenological Odderon, a $C = P = -1$ partner of the $C = P = +1$ Pomeron, could exist \cite{Nicolescu}. It was suggested
\cite{Nachtmann} that high-energy photoproduction of $C=+$ mesons, e.g.
$\pi^0$, $f_2^0(1270)$ and $a_2^0(1320)$, with nucleon excitation
would provide a clean signature for odderon exchange. In particular, it was theoretically predicted the following cross section at $\sqrt{s} = 20$ GeV for the $f_2(1270)$ meson \cite{Odderon}:
\begin{eqnarray}
\sigma(\gamma p \rightarrow f_2^0(1270) X) \approx 21\,{\mathrm{nb}}.
\label{predict}
\end{eqnarray}
whereas the the experimental results at $\sqrt{s} = 200$ GeV for $f_2(1270)$ \cite{H1f2m} is $\sigma(\gamma p \rightarrow f_2^0(1270) X) < 16$ nb at the 95$\%$ confidence level.

The model referred above is based on an approach to high-energy diffractive scattering
using functional integral techniques  and an extension of the model of the stochastic vacuum (for details, see Ref. \cite{Odderon}).  It is easily extended to Odderon exchange and gives an Odderon intercept $\alpha_{\mathrm{odd}}(0) = 1$. The nucleon and the baryon resonances are treated as quark-diquark dipole systems. The
wave functions automatically take
into account helicity flip at the particle and at the quark level and
produce the correct helicity dependence of $d\sigma/dt$ as $t \to 0$ for
Regge-pole exchange. In elastic hadron-hadron scattering the
increase of the cross sections, together with the shrinking of the
diffractive peak, can be reproduced in this model by suitable scaling of
the hadronic radii. The assumption that the same radial scaling is relevant
for the energy dependence of the Odderon contributions, leads to the
photoproduction cross sections scaling as $(\sqrt{s}/20)^{0.3}$ \cite{Odderon}.

\begin{table}[t]
\centering
\renewcommand{\arraystretch}{1.5}
\begin{tabular}{l c c }
\hline
 $f_2(1270)$  & $\sqrt{s}=7$ TeV & $\sqrt{s}=14$ TeV \\
\hline
    $\frac{d\sigma}{dy}(y=0)$  & 2.4 nb & 2.9 nb \\
    $\sigma_{tot}$ &  29 nb & 37 nb \\
  \hline
\end{tabular}
\caption{Rapidity distribution at $y=0$ and integrated cross sections for the photon-Oderon production channel at the LHC energies \cite{MVTMPRD}.}
\label{tab3}
\end{table}

In what follows we estimate the photon-Odderon contribution (meson photoproduction) to the $f_2(1270)$ exclusive production using the equivalent photon approximation. In this case, the proton-proton cross section can be written as the convolution of the probability of the proton emit a photon with the photon-nucleon cross section producing a resonance ($\gamma p \rightarrow f_2+N$):
\begin{eqnarray}
\sigma_{pp(\gamma p)\rightarrow pRp}(\sqrt{s}) = \int_{Q_{min}^2}^{Q_{max}^2}\int_{x_{min}}^1\frac{d^2n_{\gamma}}{dQ^2dx}\sigma_{\gamma p \rightarrow R}\,dQ^2dx,\nonumber \\
\,\,
\label{foto}
\end{eqnarray}
where we use the equivalent luminosity spectrum for the proton, $d^2n_{\gamma}/dQ^2dx$ (see Ref. \cite{MVTMPRD} for details). 

In Ref. \cite{Odderon} the authors have computed the photoproduction cross section of $f_2$ meson at the energy $W_{\gamma p}=20$ GeV obtaining the value 21 nb. Moreover, there it has been determined that the energy behavior for Odderon exchange in the photoproduction cross sections scales as $\sigma\,(W_{\gamma p})\propto W_{\gamma p}^{0.3}$. For the photoproduction cross section, we take a simple extrapolation based on the theoretical arguments presented above:
\begin{eqnarray}
\sigma (\gamma p \rightarrow f_2(1270)N) = \sigma(W_0)\left(\frac{W_{\gamma p}^2}{W_0^2} \right)^{0.15},
\end{eqnarray}
where $\sigma(W_0)=21$ nb and $N$ is the nucleon excitation. The energy scale $W_0=20$ GeV is considered, in which the Odderon contribution has been computed \cite{Odderon}.    

 Putting the extrapolation above in Eq. (\ref{foto}), we obtain an estimation of contribution associated to the photon-Odderon production channel. At $\sqrt{s}=20$ GeV the theoretical uncertainty was estimated to be of a factor $2$ \cite{Odderon2}. Similar trend should remain in present case. In the case considered here, the proton is required to break up. This is interesting, since currently all LHC experiments have insufficient  forward coverage, which does not allow a full reconstruction of central exclusive processes.The numerical results for the rapidity distribution at central rapidity and the integrated cross sections are presented in Table III.  Concerning the photoproduction in this channel, it should be noticed that charge asymmetry in the $\pi\pi$ mass spectrum around $f_0$ and $f_2$ mesons may signal the Pomeron-Odderon interference effects as described for instance in \cite{Pire}.

As a summary, we have investigated the central diffractive production of mesons $f_0(980)$ and $f_2(1270)$ at the energy of CERN-LHC experiments on proton-proton collisions. For the central diffraction processes we have considered two  non-perturbative Pomeron model to the meson production. In particular, the Donnachie-Landshoff Pomeron model is able to provide the cross section for $J=1,2$ meson states like $f_2(1270)$. The main predictions are the differential cross section for exclusive diffractive $f_0(980)$ meson production, $d\sigma/dy (y=0)\simeq 27$ $\mu$b at the LHC energies as an upper limit and total cross section for exclusive diffractive $f_2(1270)$ meson production, $\sigma\,(f_2)\simeq 1100$ nb.  The theoretical uncertainties are large in such cases, as discussed in text.  We have also verified the role played by the photon-Odderon production channel.  Namely, we study the implication of high-energy photoproduction of $C=+1$ mesons as $f_2(1270)$ with nucleon excitation through Odderon exchange. We found that such a contribution could be relevant if proton tagging is not imposed.
Concerning the decay channels for the mesons investigated here, the two-pion decay is the dominant one. The branching ratio for $f_2(1270)\rightarrow \pi\pi$ is about 84.8 \%  and for the $f_0(980)$ meson one has $\Gamma (\pi\pi)/[\Gamma (\pi \pi)+\Gamma (K\bar{K})]=0.75$ \cite{PDG}. Actually, this is the verified signal measured at the ALICE experiment that is able to detect the pion pair for double and no-gap events. In particular, in the double gap distribution, the $K_s^0$ and $\rho^0$ are highly suppressed while the $f_0(980)$ and $f_2(1270)$ with quantum numbers $J^{PC}=(0,2)^{++}$ are much enhanced. This enhancement for such states is evidence that the experimental double gap condition used for ALICE selects events dominated by double Pomeron exchange.


\begin{references}

\bibitem{ALICE} R. Schicker [ALICE Collaboration], arXiv:1110.3693 [hep-ex].

\bibitem{MVTMPRD} 
  M.~V.~T.~Machado,
  Phys.\ Rev.\ D {\bf 86}, 014029 (2012).


\bibitem{Land-Nacht} P.V. Landshoff and O. Nachtmann, Z. Phys. \textbf{C35},
405 (1987).

\bibitem{Bial-Land} A. Bialas and P.V. Landshoff, Phys. Lett. \textbf{B256
}, 540 (1991).

\bibitem{KKMR} A.B.~Kaidalov, V.A.~Khoze, A.D.~Martin and
M.G.~Ryskin, Eur. Phys. J. {\bf C 21}, 521 (2001).

\bibitem{KMRCHIS2} L.A. Harland-Lang, V.A. Khoze, M.G. Ryskin and W.J. Stirling, Eur. Phys. J. {\bf C 65}, 433 (2010).

\bibitem{TevCHI} T. Aaltonen {\it et al.} [CDF Collaboration], Phys. Rev. Lett. {\bf 102}, 242001 (2009).


\bibitem{KMRS} V.A.~Khoze, A.D.~Martin, M.G.~Ryskin and W.J. Stirling, Eur. Phys. J. {\bf C 35}, 211 (2004).


\bibitem{Farrar} M. B. \c{C}akir and  G. Farrar, Phys. Rev. {\bf D50}, 3268 (1994). 

\bibitem{hep_lowx} M.V.T. Machado and M. L. L. da Silva, arXiv:1111.6081 [hep-ph].

\bibitem{Yuan} F. Yuan, Phys. Let. {\bf B510}, 155 (2001).

\bibitem{DOLA} A. Donnachie, P.V. Landshoff, Nucl. Phys. B {\bf 303}, 634 (1998).

\bibitem{KKS} J.H. Kuhn, J. Kaplan and E.G.O. Safiani, Nucl. Phys. {\bf B157}, 125 (1979).

\bibitem{Barnes} E.S. Ackleh and T. Barnes, Phys. Rev. {\bf D45}, 232 (1992).

\bibitem{WA102}  A. Kirk,  Phys. Lett. \textbf{B489}, 29 (2000).


\bibitem{Nicolescu} R.F. Avila, P. Gauron and B. Nicolescu,  Eur. Phys. J. {\bf C 49}, 581 (2007).

\bibitem{Nachtmann} W. Kilian and O. Nachtmann,  Eur. Phys. J. {\bf C 5}, 317 (1998).

\bibitem{Odderon} E.R. Berger, A. Donnachie, H.G. Dosch and O. Nachtmann, Eur. Phys. J. {\bf C 14}, 673 (2000).


\bibitem{H1f2m} T. Berndt (for the H1 Collaboration), Acta Phys. Polonica B {\bf 33}, 3499 (2002). 

\bibitem{Odderon2} A. Donnachie, H.G. Dosch  and O. Nachtmann,  Eur. Phys. J. {\bf C 45}, 771 (2006).

\bibitem{Pire} P. Hagler, B. Pire, L. Szymanowski and O. Teryaev, Phys. Lett. {\bf 535}, 117 (2002); Eur. Phys. J. {\bf C 26}, 261 (2002).

\bibitem{PDG} K. Nakamura {\it et al.} (Particle Data Group), J. Phys. G: Nucl. Part. Phys. {\bf 937}, 0755021 (2010).



\end{references}
\end{document}